\RequirePackage{amsthm}

\documentclass[sn-mathphys,Numbered]{sn-jnl}

\usepackage{orcidlink}%
\usepackage{graphicx}%
\usepackage{multirow}%
\usepackage{amsmath,amssymb,amsfonts}%

\usepackage{mathrsfs}%
\usepackage[title]{appendix}%
\usepackage{xcolor}%
\usepackage{textcomp}%
\usepackage{manyfoot}%
\usepackage{booktabs}%
\usepackage{algorithm}%
\usepackage{algorithmicx}%
\usepackage{algpseudocode}%
\usepackage{listings}%

\theoremstyle{thmstyletwo}%

\theoremstyle{thmstylethree}%

\begin{document}

\title[Immersive Analysis: Enhancing Material Inspection of X-Ray Computed Tomography Datasets in Augmented Reality]{Immersive Analysis: Enhancing Material Inspection of X-Ray Computed Tomography Datasets in Augmented Reality}


\author*[1,2]{\fnm{Alexander} \sur{Gall} \orcidlink{0000-0002-3649-7368}}\email{alexander.gall@uni-passau.de}

\author[1,3]{\fnm{Anja} \sur{Heim}}

\author[1]{\fnm{Patrick} \sur{Weinberger}}

\author[1]{\fnm{Bernhard} \sur{Fröhler}}

\author[1]{\fnm{Johann} \sur{Kastner}}

\author[2,3]{\fnm{Christoph} \sur{Heinzl}}

\affil[1]{\orgdiv{Research Group Computed Tomography}, \orgname{University of Applied Sciences Upper Austria}, \orgaddress{\street{Roseggerstraße 15}, \city{Wels}, \postcode{4600}, \country{Austria}}}

\affil[2]{\orgdiv{Faculty of Computer Science and Mathematics}, \orgname{University of Passau}, \orgaddress{\street{Innstraße 43}, \city{Passau}, \postcode{94032}, \country{Germany}}}

\affil[3]{\orgdiv{Division Development Center X-ray Technology}, \orgname{Fraunhofer Institute for Integrated Circuits IIS}, \orgaddress{\street{Flugplatzstraße 75}, \city{Fürth}, \postcode{90768}, \country{Germany}}}



\abstract{This work introduces a novel Augmented Reality (AR) approach to visualize material data alongside real objects in order to facilitate detailed material analyses based on spatial non-destructive testing (NDT) data as generated in X-ray computed tomography (XCT) imaging. For this purpose, we introduce a framework that leverages the potential of AR devices, visualization and interaction techniques to seamlessly explore complex primary and secondary XCT data matched with real-world objects. The overall goal of the proposed analysis scheme is to enable researchers and analysts to inspect material properties and structures onsite and in-place. Coupling immersive visualization techniques with real physical objects allows for highly intuitive workflows in material analysis and inspection, which enables the identification of anomalies and accelerates informed decision making. As a result, this framework generates an immersive experience, which provides a more engaging and more natural analysis of material data. A case study on fiber-reinforced polymer datasets was used to validate the AR framework and its new workflow. Initial results revealed positive feedback from experts, in particular regarding improved understanding of spatial data and a more natural interaction with material samples, which may have significant potential when combined with conventional analysis systems.}




\maketitle


\section{Introduction}\label{sec1}
    In materials science, the development and continuous optimization of novel and advanced materials is essential. 
    Supported by the rapid digitalization of industry, increasing amounts of complex, heterogeneous and high-dimensional data are generated, especially in the area of non-destructive testing (NDT) \cite{Ida2019, Heinzl2019, Heinzl2017} of materials or components. 
    In order to meet the key objectives of this digital revolution (cf. Industry 4.0) regarding location independent working, real-time monitoring and visualization, quality control and process optimization, respective workflows are required for advanced data analyses. 

    Recent technologies in immersive analytics, e.g., in terms of augmented reality, offer new ways to analyze and explore large heterogeneous datasets and support users in cognitively demanding tasks. 
    Considerable progress has been made in this direction through digital support in industrial application areas such as assembly, maintenance or training \cite{Ho2022, Evangelista2020}. 
    But so far, there are only a few applications in materials characterization, quality control and metrology, that employ new immersive forms for the analysis of heterogeneous materials science data. 
    So, an increasing need regarding immersive visualization systems becomes obvious. Such systems need to facilitate intuitive analyses of complex data from advanced imaging techniques such as X-ray computed tomography (XCT), e.g., to investigate defects onsite at the actual component or to explore the internal micro-structure of materials in a purely virtual domain. 
    AR can be used via head-mounted displays (HMD) to conduct analyses anywhere. Augmented Reality (AR) has the ability to overlay virtual data on real objects and link them. Therefore, inspections and analyses are no longer limited to office settings but can be conducted on-site, making the process more flexible and efficient. Since AR utilizes the real environment, it can provide contextual information, bridging the gap between the users and their environment. For example, automatically recognizing samples stored in a database allows for a highly intuitive and simple data access for subsequent analyses. Additionally, AR HMDs provide the opportunity to stay aware of the local surroundings while maintaining both hands free for other activities, which is a significant advantage in an industrial context \cite{Ens2021a}. In addition, AR HMDs offer the possibility to use other visible systems without restrictions, thus expanding the possible range of available functions. Productivity and accuracy is improved making use of immersive complementary interfaces that supplement the workflow in a meaningful way \cite{Ho2022, Zagermann2022}. Research also indicates that AR can enhance understanding and comprehension of data visualization, achieving the same level of accuracy as 2D visualizations while providing a more immersive user experience \cite{Schroeder2020, Billinghurst2018}. Specifically, AR's capability to enable users to walk through and zoom into visualizations intuitively improves the exploration and analysis process \cite{Wang2020a}.
        
    Our work presents a framework for immersive analysis of complex materials science data using AR. We explore the positive impact of embodied navigation and interaction, as well as the natural analysis of three-dimensional volumes and their physical counterparts on NDT data. To our knowledge, this is the first framework enabling material experts an AR analysis of spatial and abstract NDT data. We tested the potential benefits of the new workflow in a user study with experts in the field and identified opportunities for future immersive analysis in this area.

\section{Related Work}\label{sec:related-work}

    Immersive analytics as a scientific field has experienced significant growth since its origins around 2015, which is confirmed by many research activities in both AR and Virtual Reality (VR) \cite{Marriott2018}. Industry 4.0, also known as the forth industrial revolution, reflects the growing interest in this rapidly developing area regarding the analysis of large heterogeneous data. 
    
    \subsection{Opportunities of Immersive Analytics with respect to NDT Analyses}\label{sec:opportunities}

        Over the last years researchers discovered promising opportunities of immersive analytics in various application areas \cite{Chandler2015, Fonnet2019, Ens2021a, Kraus2022}. It aims to improve data understanding and decision making by bridging the gap between the users and their data through more natural, engaging, and immersive analysis. Although immersive analytics is not targeting a specific technology, a trend leaning towards from AR and VR technologies is clearly visible in novel analysis concepts. Particularly in the context of Industry 4.0, AR acts as a key enabling technology, offering increased productivity, accuracy, and autonomy in the quality sector \cite{Evangelista2020, Ho2022}. For materials science, a field of research characterized by complex spatial structures and large amounts of high-dimensional data, immersive analytics offers a significant potential for transforming tasks such as materials characterization, quality control, and metrology. For these tasks, technologies such as AR offer new ways of engaging with data analysis. This may significantly change how we perform and interact with data in future as outlined in the following paragraphs:

        \begin{itemize}
    
        \item \textbf{Embodiment:} 
        Embodied data exploration enables users to employ intuitive (body) movements when exploring complex and large data. Avoiding traditional mouse and keyboard inputs and replacing these with embodied navigation, gestures, and tangible interactions, the cognitive load can be considerably reduced by making exploration feel more natural \cite{Bueschel2018}. 
        
        \item \textbf{Spatial immersion:} 
        An almost unlimited, three-dimensional (virtual) workspace for analyses may be utilized in immersive analytics settings. The immersive workspace provides experts with perceptually convincing renderings of spatial data, and the ability to organize data visualizations in the environment and along their workflow \cite{Gall2021}. 
    
        \item \textbf{Using depth \& non flat surfaces: }
        By utilizing depth information, immersive devices enable an improved perception of spatial objects through stereoscopic rendering and the display of non-planar surfaces. Furthermore, exploiting the third dimension enables the development of new three-dimensional visualization techniques. This allows experts to virtually "step inside" the micro-structure of a material, providing a unique perspective on the material structure, which in consequence can lead to a deeper understanding \cite{Marriott2018a}.
    
        \item \textbf{Arranging multiple views:} 
        Additional information may be provided in separate views (2D sectional views, density plots of material parameters etc.), which can be arranged in different layouts in the environment. This improves the analysis by enabling experts to use the entire surrounding. AR enables users to enhance current workflows, e.g. on traditional desktop systems, by complementary interfaces providing almost infinite virtual spaces for respective analyses \cite{Zagermann2022, Marriott2018a}.
        
        \item \textbf{Engagement:} 
        Working within immersive environments offers users new ways to interact with their data and a substantially increased sense of presence, allowing them to better engage in their data analysis tasks. As a result engagement can create a enjoyable state, also known as flow, which strongly affects the experience of users and degree of productivity \cite{Marriott2018a}.
       
        \item \textbf{Situated analytics:}  
        In situated analytics the physical world is augmented with additional information, which is displayed aside the real object or overlaid to add context-sensitive information. As a result on-site and in-place data analytics are facilitated, by directly interacting with the object of interest. Using mobile immersive devices, situated analytics thus becomes location independent \cite{Willett2017, Ens2021a, Bressa2022}. In industrial application scenarios see-through AR HMDs provide additional advantages: Being able to observe the surrounding environment, workers can utilize devices even in dangerous workplaces. They do not have to share their attention regarding reality and augmentation and still have their hands free for interaction or handling samples \cite{Ens2021a}. 
        
        \item \textbf{Collaboration:} 
        Immersive technologies offer the opportunity to collaborate. The combination of embodied data exploration, spatial immersion, and shared environments facilitates socially engaged co-located and remote collaboration. In an industrial setting, this approach can eliminate the requirement of on-site experts for inspection tasks \cite{Gall2022a}. The collaborative analysis of multi-dimensional data also benefits from novel input modalities and new capabilities provided by an almost infinite immersive workspace and virtual avatars \cite{Billinghurst2018}.
        
        \item  \textbf{Visualising abstract data with a spatial embedding:} 
        Finally immersive analytics excels when it comes to the analysis of spatial data with additional abstract information, as it is common in material analysis. A promising concept are Information-Rich Virtual Environments (IRVEs) \cite{Polys2011, Marriott2018a, Skarbez2019} where inherently spatial data is harmoniously integrated with additional abstract data. This enriches the user's understanding of the dataset, creating an even more comprehensive and seamless user experience. As a result, in an AR application, experts can perform more natural inspections by interacting directly with the material.
               
        \end{itemize} 

        Considering all these opportunities, we have come to the conclusion that immersive analytics is a promising research direction for the field of materials science, and AR as a promising technology for industrial applications. 

    \subsection{Applications of Immersive Analytics in NDT}
   
        Our focus regarding applications is on immersive analytics systems that employ spatial data, particularly within the domain of material science and that facilitate an analysis. There are few immersive analytics systems available analyzing volumetric data as well as high-dimensional data, and they are divided between AR- and VR-focused systems. However, systems that allow for the material characterization of the generated NDT data by means of immersive techniques are not widely available. This highlights a significant gap in literature. A large proportion of applications in the industrial environment involves assembly, maintenance, and training tasks. Such tasks are not relevant to our research focus. Here, we rather present the most relevant work for AR and VR based analysis and visualization of NDT data.
        
        Wang et al.~ \cite{Wang2020b} proposed the manipulation of CT datasets and their transfer function in VR. Their system allows tweaking various rendering parameters using hand gestures, but it does not offer any analysis options. Tadeja et al. \cite{Tadeja2020} use VR for creating an aerospace design environment for digital twinning. Their tool allows for loading CAD geometries of aircraft components and respective exploration of their performance parameters through 3D scatterplots. ImNDT by Gall et al.~\cite{Gall2021} offers material experts an immersive workspace for the analysis of multi-dimensional data from NDT. The authors developed visual metaphors and interaction techniques for their VR system, which can only be applied to secondary data and do not allow a visualization of voxel data. Like in any other VR system, users are also here isolated from the real world. Gall et al.~\cite{Gall2022} extended the ImNDT system to allow for Cross-Virtuality-Analysis (XVA). They added an AR view mode for the VR user to enable collaboration with users of a large touch screen. This study found that XVA as form of analysis is beneficial, but suffers in the presented application from the poor resolution of the used VR cameras.

        Schickert et al.~\cite{Schickert2018} demonstrated the use of AR for material inspection. The authors developed an application for a tablets to superimpose CAD models extracted from indications of ultrasound and radar images onto a concrete specimen. The tool positions CAD models in the location of a QR code, but does not provide the ability to modify the geometry or analyze additional secondary data. The system introduced by Ferraguti et al.~\cite{Ferraguti2019} supports workers inspecting polished surfaces by projecting measurement data directly onto the component's surface for intuitive quality evaluation. While visual analysis is possible, the system does not facilitate the exploration of additional data that is required for our material characterization. Corsican Twin by Prouzeau et al.~\cite{Prouzeau2020} combines VR and AR devices to create an immersive authoring tool. It allows users to annotate the virtual replica of a real-world environment in VR, and displays the charts or CAD models to AR users in the real-world environment through placed QR codes. For material characterization, the approach does not provide opportunities to explore large amounts of high-dimensional data in addition to the real samples.

\section{Design \& Methods}\label{sec:design}

    To outline the design requirements and the structure of the proposed immersive analytics system, we first discuss the available datasets, their representation, the proposed architecture of the framework and finally the implemented interaction and visualization techniques. The proposed framework serves as a prototype to enable immersive analysis of materials, which works independent from traditional desktop settings. For this prototype, we are building on promising results from previous work in immersive analysis of spatial data \cite{Gall2021b, Gall2021, Gall2022, Tadeja2020, Ferraguti2019, Schickert2018}. 

    \subsection{Material Data} \label{subsec:materials}
        
        The NDT data used is mainly obtained from composite materials, more precisely from fiber-reinforced polymers. The primary data is obtained by using various XCT imaging devices and consists of 3D reconstructed volume datasets, stored as a binary file (*.raw). Multidimensional data, i.e., secondary data, is derived from the generated primary data. The secondary data used in this framework was derived via the fiber characterization pipeline of Salaberger et al.~\cite{Salaberger2011}. In this process a Hessian matrix is used for finding the medial axis of the individual fibers after applying Gaussian blurring for noise reduction. The result is a multi-dimensional table of attributes featuring the individual characteristics for each fiber, such as its length, diameter, start/end point, and volume, among others. This information is stored in a comma-separated file (.csv).

    \subsection{Framework Structure}

        Given the current lack of appropriate AR systems for materials characterization, indicated in \autoref{sec:related-work}, we developed a new framework to seamlessly integrate complex primary and secondary XCT data with real-world objects. This enables researchers and analysts to inspect material properties and structures in place, offering them a virtual workspace. Using immersive visualization techniques coupled with the real physical objects, users can adopt novel, highly intuitive workflows for material analysis and inspection, which helps to identify anomalies easier enabling fast informed decision-making.

        The proposed framework is based on the Unity engine (Version 2021.3) \cite{Unity} for rendering immersive environments. Unity is used because of its wide variety of capabilities and compatibility with numerous immersive devices and platforms. For this reason it has been adopted for a lot of immersive applications and also serves as an effective framework for InfoVis applications, demonstrating its versatility across different areas \cite{Wagner2016, Fonnet2019, de_Souza_Cardoso2020}. Using Unity is particularly advantageous, as it is offering the possibility to extend the framework to other virtual reality, augmented reality or mixed reality devices. To enable the tracking of objects, we use the Vuforia Engine \cite{Vuforia} which enables the robust detection and tracking of multiple targets, such as images and objects. The integration with Unity is straight forwards and the interaction with real-world objects simple. Based on the reasons mentioned in \autoref{sec:opportunities}, we are using a see-through AR HMD as immersive technology. Specifically, we have opted for the Microsoft Hololens 2 \cite{Hololens2}. The device provides advanced spatial recognition and eye-tracking, which enhances interaction with objects and virtual representations. To facilitate further development, the framework will be released as open source on GitHub.  
    
\begin{figure}[tbh] 
        		\centering
        			\includegraphics[scale=0.99]{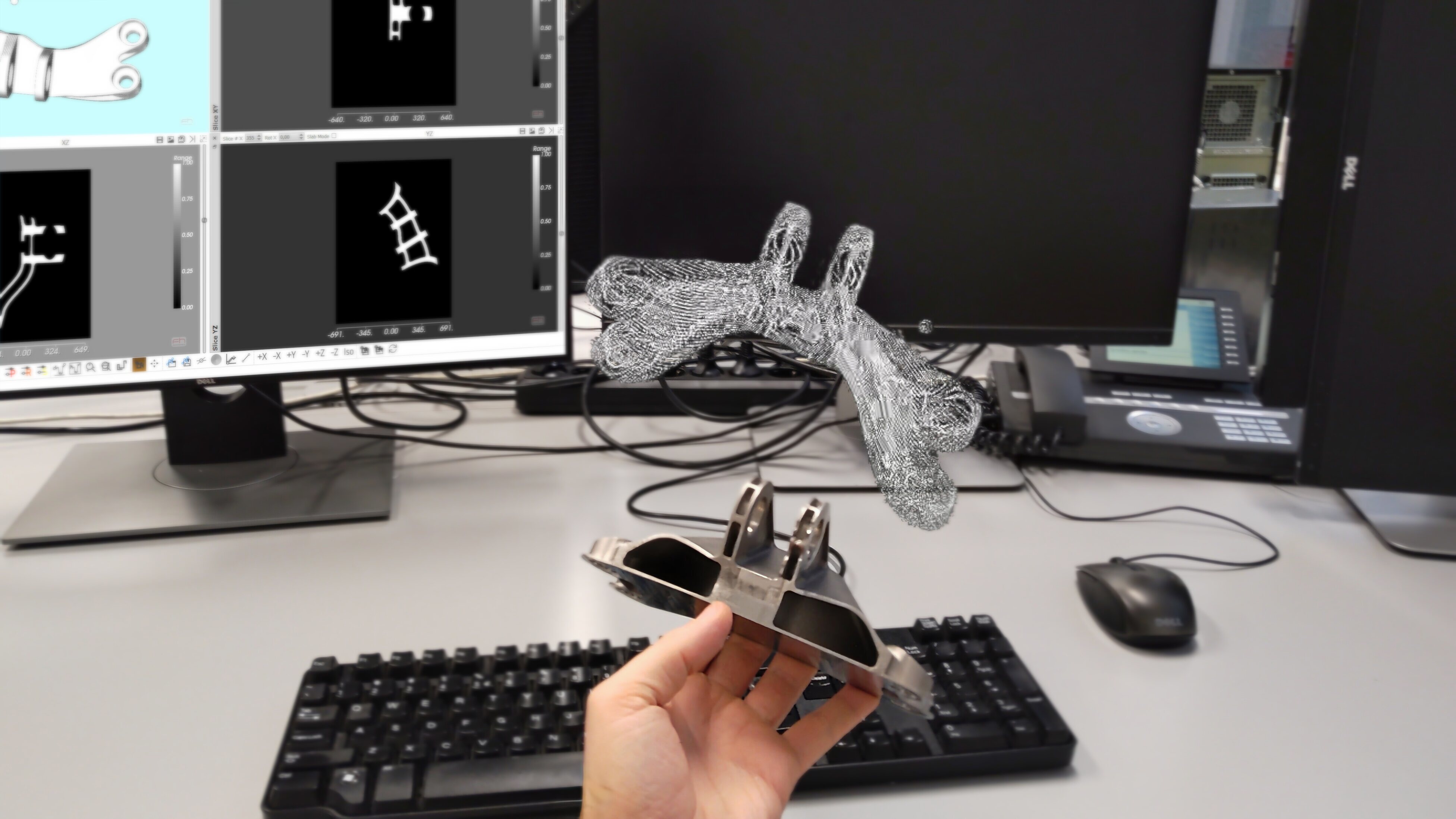}		
        		\caption{The AR HMD uses the stored shape to detect the material. The image of the sample captured by XCT is displayed alongside the physical sample and moves in sync with it. A traditional analysis can still be carried out via monitors in the background.}
        		\label{fig:SpatialRecognition}
        	\end{figure}

        Our visualization framework was primarily designed as an extensible platform for performing immersive analyses on XCT data. The developments are specifically targeted but not limited to the domain of materials science with regard to fiber characterization. Our system enables the analysis of both primary and secondary derived datasets. Users may select and load XCT datasets manually using a file dialog or by physical object recognition using the cameras on the AR HMD. The latter can be performed in two different ways, which are image target recognition or shape recognition. Using image target recognition, the system detects and tracks images attached to a sample that were previously stored and linked to a specific XCT dataset. The shape recognition uses existing shape data, such as a CAD model or a surface mesh created via photogrammetry, to detect the sample and then load and display the corresponding XCT dataset. \autoref{fig:SpatialRecognition} illustrates the detection process using a CAD model. The primary dataset is linked to the real object, which acts as a proxy for interactions. The XCT dataset is displayed using various rendering techniques, such as direct volume rendering, maximum intensity projection, or similar.
        The user can load secondary data, such as lists of attributes from features of interests such as defects or other spatial objects. This information is extracted from a multi-dimensional table of characteristics stored as .csv file. In the case of fiber reinforced polymers this includes information on their length, diameter, and orientation. If spatial information, i.e., start/end points of fibers, is available in the derived data an abstracted representation can be used for further exploration. In our application, we use a model-based surface representation to depict the fibers as cylinders. This visualization is created using the fibers' start and end points, along with their radii. This approach allows for a simplified and less cluttered visual representation of the volume, facilitating the exploration of larger datasets. 
        Abstract data is also highly important for a detailed analysis of materials. For records of secondary data in the .csv file, which cannot be spatially interpreted, such as stress tensor fields and statistics of characteristics, various charts as histograms or scatter plots may be generated. The visualization techniques and interaction paradigms will be explained in more detail in \autoref{subsec:interaction-vis}. Since the framework runs as a standalone application on the Hololens, it does not require a PC in an office or the factory hall. Therefore, location independent analysis can be performed anywhere at any place and any time. To extend the analysis, additional systems, such as traditional desktop-based analysis systems, can be included in the inspection (see \autoref{fig:SpatialRecognition}). This is possible because see-through AR allows the natural environment to be retained, so the application can be used as a complementary interface to an existing traditional system.

    \subsection{Visualization Techniques and Interaction Paradigms} \label{subsec:interaction-vis}     

        After identifying a sample, the corresponding XCT volume and abstract representation (depending on what is available) are displayed alongside the physical material. Users interact with spatial representations using hand gestures. By grabbing a representation with two hands and pulling the hands apart or together, a zooming metaphor is applied. Similarly, the visualizations are rotated and positioned by intuitive gestures. Furthermore, natural interaction with volumes is also possible by just interacting with the physical sample at hand. This results in a more natural, embodied inspection of the material.         
        
\begin{figure}[tbh] 
        		\centering
        			\includegraphics[scale=0.99]{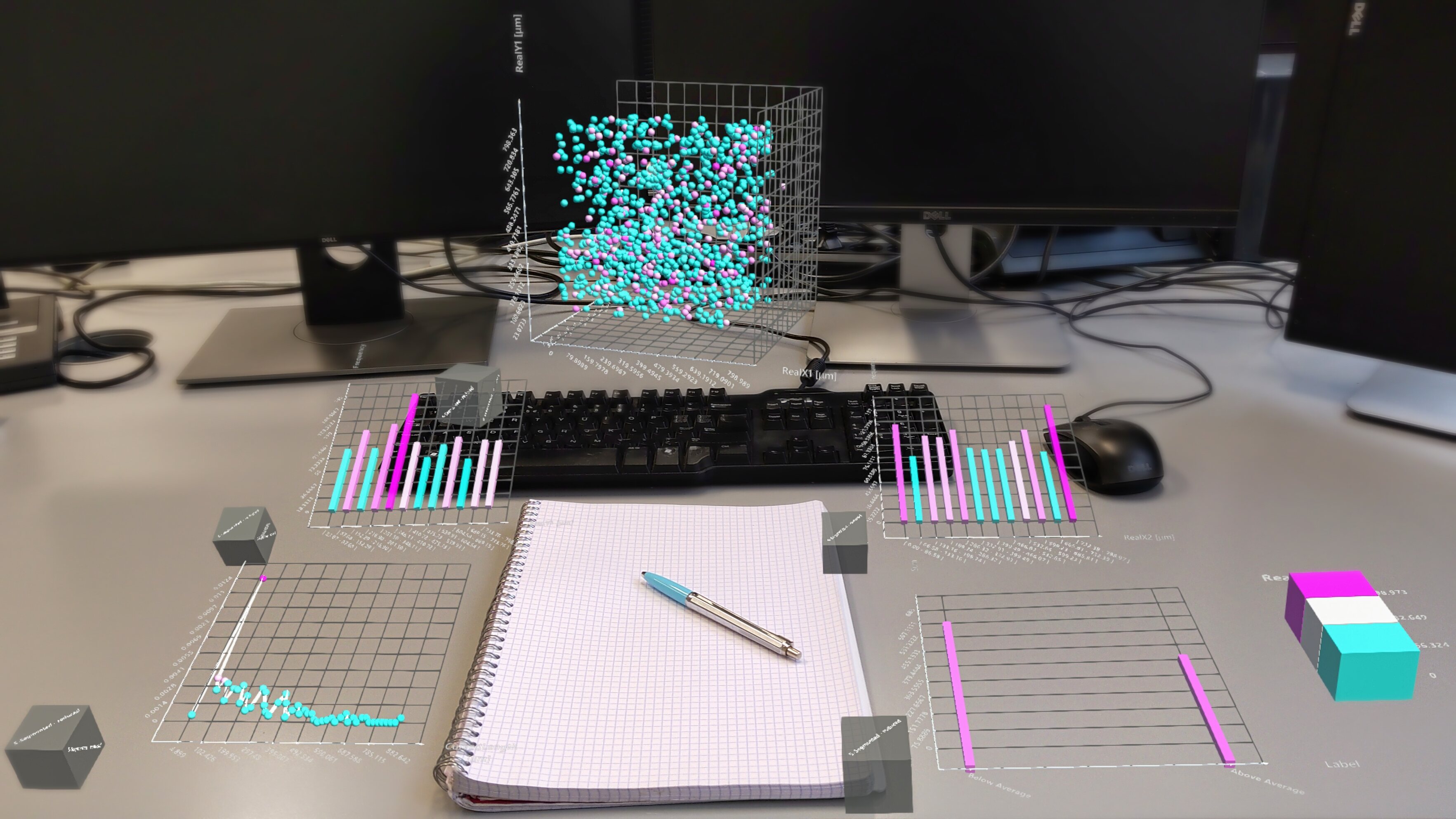}		
        		\caption{Visualization techniques for analyzing abstract data: The user in this example utilizes histograms, bar charts, a distribution plot, and a three-dimensional scatterplot to display different fiber characteristics, and has arranged these visualizations in the workspace.}
        		\label{fig:AbstractVis}
        	\end{figure}

        The framework offers various visualization techniques to analyze the secondary data, i.e., abstract data, in addition to the spatial data. These include histograms, 3D scatterplots, bar charts, and density plots. \autoref{fig:AbstractVis} shows a collection of abstract visualization techniques arranged in an office setting. These charts may be created at arbitrary places in the world and are interactive during the analysis. Interactions with the virtual representations are again carried out using hand gestures through grabbing a handle in the form of a cube placed at the lower left side of each representation. The gestures applied to the handle allow for zooming, rotating, or similar interactions using a single or both hands. Visualizations may be generated for all attributes present in the secondary dataset. The histogram visualization displays the frequency distribution of a selected attribute, allowing for a quick understanding of the most common value ranges. The 3D scatterplot reveals correlations between three attributes and provides a deeper understanding of multi-dimensional relationships. The bar chart allows experts to compare various abstract data, enhancing their understanding of differences or similarities. The distribution plot provides experts with an analysis of the shape of the distribution of an attribute for understanding the variability, central tendency, and the presence of potential outliers in the data. The visual representations provide users with a quick overview of the material's characteristics and structure, allowing them to conduct analysis in an immersive environment.
    
\section{Results \& Discussion} 
    
    The presented framework is a starting point facilitating further research in AR based material analysis. For this reason our initial studies are mainly focused on qualitative evaluation. The aim was to find out if and how immersion helps simplifying complex analysis tasks.
    
\begin{figure}[tbh] 
    		\centering
    			\includegraphics[scale=0.99]{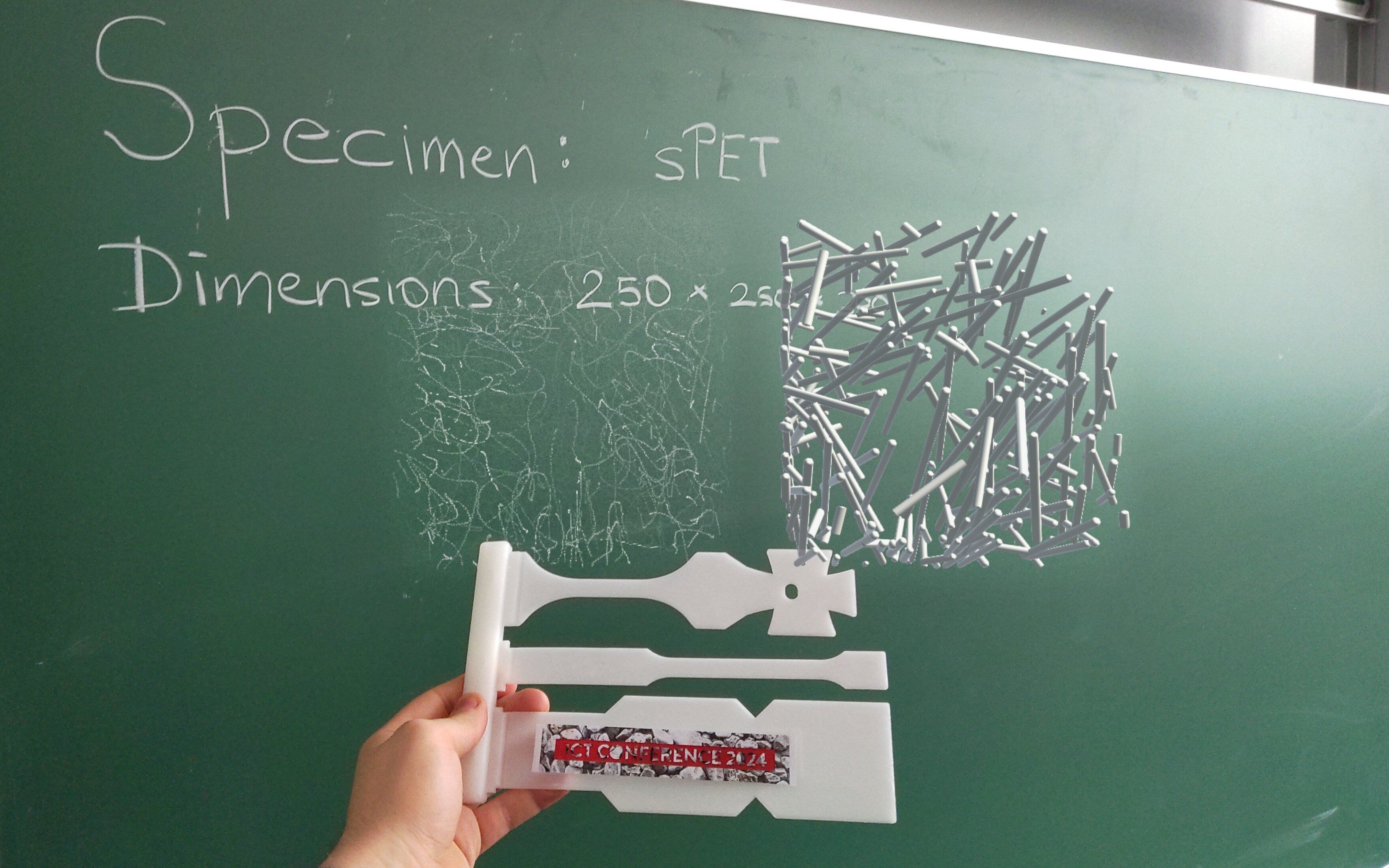}		
    		\caption{ The material is recognized by tracking the image target, which automatically loads the associated data. The XCT scan and model-based surface representation are both displayed and synchronized with the position and orientation of the material sample.
            }
    		\label{fig:UseCase01}
    	\end{figure}

    We conducted a user study with materials science experts and experts in visualization. Using a cognitive walk-through as evaluation method, we were able to gather valuable feedback on the interaction design, the visual metaphors introduced and the workflow the experts applied. As use case for this evaluation we selected a material system with polyethylene terephthalate fibers in a polypropylene matrix material. The primary dataset of this material system was obtained from a respective CT scan and secondary data was subsequently derived by the experts through the fiber characterization pipeline outlined in \autoref{subsec:materials}. The investigated sample was cut out of a standard multi-purpose test specimen manufactured by injection molding. The CT dataset yielded a size of $250 \times 250 \times 300$ voxels and contains 214 fibers, for each of which 20 distinct characteristics were computed. The validation of the data is not within the scope of our study. As the material is very thin and features a very homogeneous texture, two image targets were used for tracking. The image target on the front of the sample is visible in \autoref{fig:UseCase01}. As most of the study participants possess a background in analyzing fiber reinforced polymers, an exploration task was defined to ensure the expert's workflow was in no way constrained. This enabled us to assess how the experts utilize our tool and identify any issues with it. Four participants were selected, including two experts in visualization and two specialists in materials science and computer tomography. Participants were given the freedom to use the application while sitting or standing as well as to ask about controls. To gather qualitative insights, they were instructed to use the 'Thinking Aloud' technique during the testing phase, expressing their planned actions and any conclusions they drew. Prior to conducting the study, participants received instructions on the general controls and the available representations in the application. 

\begin{figure}[tbh] 
        \centering
            \includegraphics[scale=0.99]{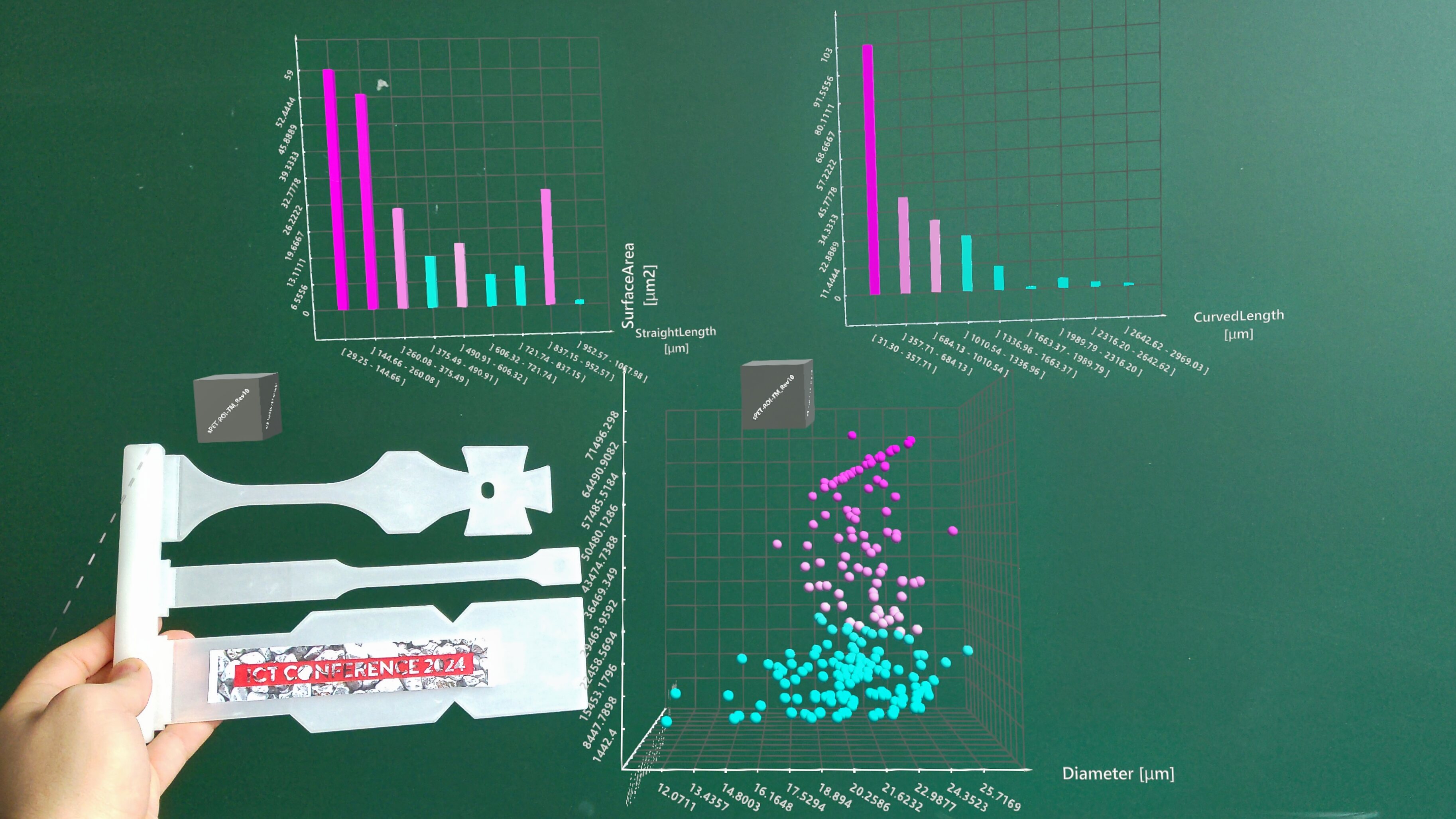}		
        \caption{After recognizing the dataset, different visualization techniques can be utilized. In this case, two histograms (top) and a 3D scatterplot (bottom) are used to analyze, e.g., the lengths and diameters of the fibers.}
        \label{fig:UseCase02}
    \end{figure}

    For our study we used the Microsoft Hololens 2 see-through AR HMD as immersive device to run the application. The analysis starts with fitting and setting up the HMD. Then, the participants pick up the prepared material sample from the table to inspect it. After the detection phase of the sample, the relevant data stored on the Hololens is loaded automatically. The user is presented with the volume rendering of the XCT dataset and its model-based surface representation, both hovering over the sample (see \autoref{fig:UseCase01}). The virtual representations are manipulated in space using the physical sample. Additionally, pre-selected data visualizations for the secondary data are displayed next to the sample in the environment. In our use case, these are two histograms for different fiber lengths and a three-dimensional scatterplot encoding various spatial dimensions, which can be seen in \autoref{fig:UseCase02}. The first histogram visualizes the distribution of straight fiber lengths (the distance between the start and end points of the fiber), while the second histogram shows the distribution of actual lengths considering the curvature of each fiber. This approach enables the identification of clusters related to specific fiber lengths and curvature. The scatter plot shows the correlation of the fibers based on their diameter, area and length. This visualization enables users to draw conclusions regarding whether the fibers tend to elongate or thicken more, affecting their surface area. The representations of abstract data in the environment can be freely moved and enlarged by experts to adapt to their current investigation phase or question.

    During the exploration phase, which on average lasted 25-minutes, experts were able to inspect materials and test the application's functionality. Valuable feedback on the benefits of an immersive exploration of material datasets was provided by subsequent discussions with the experts. This feedback yielded that the presented framework seems to be promising for various material characterization tasks, with a significant advantage being the ability to interact directly with the material sample. The synchronization of translations and rotations between the physical samples and virtual representations creates a natural interaction, eliminating the need for traditional mouse and keyboard operations. Additionally, as the data is automatically loaded when the user starts interacting with the real sample, no further input is required, and the analysis is intuitively initiated. Some participants with prior experience in VR systems mentioned that the AR device is much lighter and more comfortable to use for longer data exploration. Additionally, it requires no cables and does not isolate them from their office environment and colleagues. The experts however also identified some disadvantages: The analysis was challenging in bright environments due to low color fidelity, rendering the distinction of some representations from bright backgrounds difficult. Further comments referred to current hardware limitations of AR devices. Arguably, the field of view (FoV) is rather small, which becomes particularly obvious when enlarging volumetric data. In this case it is impossible to see the whole object within the available FoV. This requires more head movement, which can lead to fatigue. Finally, some experts expressed interest in analyzing larger primary datasets exceeding 500 megabytes. The use of the Hololens 2 in standalone mode with active object tracking and visible rendering leads to performance bottlenecks in this case and an interactive analysis without delay or unstable tracking can no longer be guaranteed.
    
    Experts finally suggest that the system has significant potential for analyzing samples in combination with conventional analysis systems. Incorporating additional visualization techniques can enable more detailed analysis by highlighting defects in volume representations or providing insights for abstract 4D data.

\section{Conclusion \& Future Work}
    Our research resulted in the development of a novel, innovative augmented reality framework. The new system allows for immersive analysis of materials data through AR HMDs and represents a significant improvement over existing methods. A central element of our proposed framework is the implementation of situated analytics, fostering context-aware visualization and interactions. This approach leverages users' gaze direction at real specimens for data loading, enhancing usability and creating a tangible link between real material samples and their virtual renderings. The intuitive recognition of real-world material samples and the control of virtual renderings based on them provide a new level of interaction for experts. By leveraging natural depth perception and embodied navigation and interaction, it enhances the overall orientation and shape understanding of three-dimensional volume analysis. By utilizing arrangeable charts, our system provides users with an immersive, customizable, and flexible workspace that supports the analysis of abstract data. Feedback from our user study with materials specialists validates the utility and usability of our tool, marking a significant step forward in this domain. The insights gained provide researchers with innovative opportunities for future exploration in this important field.

    Through close collaboration with domain experts, we identified key areas for future enhancements to our framework. These range from advanced techniques for defect visualization, integration of additional modalities, superposition of scans on real objects, and novel visual representations for time varying data. The analysis of time-varying data is another crucial step in materials characterization. Adding novel methods to inspect primary and secondary data will facilitate understanding of changes during in-situ testing. Improvements to rendering algorithms and data compression to handle larger datasets are also of interest, alongside the potential benefits of future hardware improvements for immersive analysis.

\backmatter






\section*{Declarations}



\subsection{Funding}

The research leading to these results has received funding by research subsidies granted by the government of Upper Austria within the program line "Dissertationsprogramm der FH OÖ", grant no. 881298 "AugmeNDT" and grant no. 881309 "COMPARE", as well as “XPlain”, grant no. 895981, as well as the FTI project "X-Pro". The research has also been supported by the European Regional Development Fund in frame of the project "PEMOWE" (BA0100107) in the Interreg Bayern-Osterreich 2021-2027 programme.

\subsection{Competing interests}

The authors have no conflicts of interest to declare that are relevant to the content of this article.




\subsection{Author contributions}

All authors contributed to the conception and overall application design. The implementation was done by Alexander Gall. The main manuscript was written by Alexander Gall, with contributions and comments on previous versions of the manuscript from all authors. All authors read and approved the final manuscript.

\subsection{Code availability} 
    The framework (code, documentation), and potential sample data will be published as open source on GitHub (\href{https://github.com/GallAlex}{https://github.com/GallAlex} and \href{https://github.com/3dct}{https://github.com/3dct}).

\bibliography{sn-article}

\end{document}